\documentstyle[epsfig,12pt]{article}

\begin{document}

\hoffset = -1truecm \voffset = -2truecm \baselineskip = 10 mm

\title{\bf The gluon condensation at high energy hadron collisions }
\author{
{\bf Wei Zhu}$^a$ and {\bf Jiangshan Lan}$^b$
\\
\normalsize $^a$Department of Physics, East China Normal University,
Shanghai 200241, P.R. China \\
\normalsize $^b$Institute of Modern Physics, Chinese Academy of
Sciences, Lanzhou 730000,
 P.R. China\\
}

\date{}

\newpage

\maketitle

\vskip 3truecm

\begin{abstract}

     We report that the saturation/CGC model
of gluon distribution is unstable under action of the chaotic
solution in a nonlinear QCD evolution equation, and it evolves to
the distribution with a sharp peak at the critical momentum. We find
that this gluon condensation is caused by a new kind of
shadowing-antishadowing effects, and it leads to a series of
unexpected effects in high energy hadron collisions including
astrophysical events. For example, the extremely intense
fluctuations in the transverse-momentum and rapidity distributions
of the gluon jets present the gluon-jet bursts; a sudden increase of
the proton-proton cross sections may fill the GZK suppression; the
blocking QCD evolution will restrict the maximum available energy of
the hadron-hadron colliders.

\end{abstract}

{\bf keywords}: Gluon condensation; Unstable CGC; Gluon-jet bursts;
GZK puzzle; Blocking QCD evolution; Maximum available energy in
hadron-hadron collider

{\bf PACS numbers}: 12.38.-t; 14.70.Dj; 05.45.-a

\newpage
\begin{center}
\section{Introduction}
\end{center}

    The planning of high-energy proton-proton colliders, such as very large hadron collider (VLHC) [1] and the
upgrade in a circular $e^+e^-$ collider (SppC) [2] will provide a
nice opportunity to discover new phenomena of nature.  The hadron
collider with the center-of-mass energy of hundred $TeV$ order may
probe the parton distribution functions (PDFs) in several currently
unexplored kinematical regions. In such ultra low-$x$ region, the
PDFs maybe beyond our expectations. Therefore a new exploration of
the PDFs in the proton is necessary for any future higher-energy
hadron colliders.

    The gluon density in nucleon grows with decreasing Bjoeken variable $x$ (or increasing energy
$\sqrt{s}$) according to the linear  DGLAP
Dokshitzer-Gribov-Lipatov-Altarelli-Parisi (DGLAP) [3,4] and
Balitsky-Fadin-Kuraev-Lipatov (BFKL) [5] equations, where the
correlations among the initial gluons are neglected. At a
characteristic saturation momentum $Q_s(x)$, the nonlinear
recombination of the gluons becomes important and leads to an
eventual saturation of parton densities [6]. This state is specified
as the color glass condensate (CGC) [7], where ``condensate" implies
the maximum occupation number of gluons is $\sim 1/\alpha_s$,
although it lacks a characteristic sharp peak in the momentum
distribution.

    Recently, Zhu, Shen and Ruan proposed a modified BFKL equation in [8,9]
(see Eq. (2.1)), where the nonlinear evolution kernels are
constructed by the self-interaction of gluons as similar to the
Balitsky-Kovchegov (BK) equation [10] and
Jalilian-Marian-Iancu-McLerran-Weigert-Leonidov-Kovner (JIMWLK)
equation [11], but the former keeps the nonlinear BFKL-singular
structure. Using the available saturation models as input, the new
evolution equation presents the chaos solution with positive
Lyapunov exponents [12], and it predicts a new kind of shadowing
caused by chaos, which stops the QCD evolution after a critical
small $x_c$. This unexpected result implies that the predicted
saturation state by the BK/JIMWLK dynamics is unstable at the small
$x$ range.

    In this work, we study continually the properties of this new evolution
equation. We report that chaos in this equation converges gluons to
a state at a critical momentum. This distribution with a stable
sharp peak indicates that it is the gluon condensation (see Figs. 1
and 2). We present the evolution process from a saturated input to
the gluon condensed state in Sec. 2. We find that the chaotic
oscillations of the gluon density raise both the strong negative and
positive nonlinear corrections. The former shadows the grownup of
the gluon density, while the later is the antishadowing effect. The
antishadowing as a positive feedback process increases rapidly the
gluon density. Thus, we observed the gluon condensation at the
critical momentum $(x_c,k_c)$.

    The sharp peak in the gluon distribution is higher than the normal distribution by
several orders of magnitude due to a lot of gluons accumulated in a
narrow momentum space. The gluon condensation should appear
significant effects in the hadron processes (see Sec. 3), provided
the position of $x_c$ inters the observable range at high energy.
Obviously, we have not yet found any signals of the above mentioned
gluon condensation even at the proton-proton collider
with $E_{CM}=14~TeV$ in LHC. Therefore, we turn to the ultra high energy cosmic rays
(UHECRs) in Sec. 4, where the proton energy is much larger then that
in the accelerators. We find that a sudden increase of the
proton-proton cross section may fill the Greisen-Zatsepin-Kuzmin
(GZK) suppression [13]. Using this result we estimate the value of
$x_c$.

   Based on the results of Sec. 4, we predict the possible experimental observations of the gluon
condensation in the future high-energy proton-proton colliders in
Sec. 5. We find that the gluon condensation may bring the big
fluctuations of the gluon jets at the high-energy proton-proton
collider. The gluons in every sub-jet are monochromatic and
coherent, we call them as the gluon-jet bursts. Such intensive gluon
field provides an ideal laboratory to study QCD at the extreme
conditions. We should pay attention to the big effects of the
gluon-jet bursts when planning the next high energy hadron colliders
and the detectors. We also predict a maximum available energy of the
hadron colliders due to the blocking QCD evolution. Finally, we
discuss the reasonableness of the gluon condensation in Eq. (2.1)
from some general considerations in Sec. 6. A summary is given in
Sec. 7.

\newpage
\begin{center}
\section{The gluon condensation caused by chaos}
\end{center}

    A modified BFKL equation for the unintegrated gluon distribution $F(x,k_T^2)$ at the leading logarithmic
$(LL(1/x))$ approximation is [9]

$$-x\frac{\partial F(x,k_T^2)}{\partial x}$$
$$=\frac{3\alpha_{s}\underline{k}^2}{\pi}\int_{\underline{k}^2_0}^{\infty} \frac{d
k'^2_T}{k_T'^2}\left\{\frac{F(x,k_T'^2)-F(x,k_T^2)} {\vert
k_T'^2-k_T^2\vert}+\frac{F(x,k_T^2)}{\sqrt{k_T^4+4k_T'^4}}\right\}$$
$$-\frac{81}{16}\frac{\alpha_s^2}{\pi R^2_N}\int_{k^2_0}^{\infty} \frac{d
k_T'^2}{k_T'^2}\left\{\frac{k_T^2F^2(x,k_T'^2)-k_T'^2F^2(x,k_T^2)}
{k_T'^2\vert
k_T'^2-k_T^2\vert}+\frac{F^2(x,k_T^2)}{\sqrt{k_T^4+4k_T'^4}}\right\},
\eqno(2.1)$$where we used $F(x/2,k_T^2)\simeq F(x,k_T^2)$ near the
saturation scale; the value of $R_N=4GeV^{-1}$ is fixed by fitting
the available experimental data about the proton structure function.
The singular structure both in the linear and nonlinear evolution
kernels corresponds to the random evolution in the $k_T$-space,
where $k_T^2-k_T'^2$ may across over zero. This is a general
requirement of the logarithmic ($1/x$) resummation.

     In this section we study the properties of Eq. (2.1) using the Golec-Biernat and Wusthoff (GBW)
saturation model [14] as the input at $x_0$

$${\mathcal{F}}_{GBW}(x,k_T^2)=\frac{3\sigma_0}{4\pi^2\overline{\alpha}_s}R^2_0(x)k_T^2
\exp(-R^2_0(x)k_T^2), \eqno(2.2)$$where $\sigma_0=29.12~mb$,
$x_0=4\times 10^{-5}$, $\lambda=0.277$,
$R_0(x)=(x/x_0)^{\lambda/2}/Q_s$ and $Q_s=1~GeV$;
${\mathcal{F}}\equiv F/k_T^2$ and the parameter
$\overline{\alpha}_s$ is fixed as $\overline{\alpha}_s=0.2$. Note
that in the calculation we take $F(x,k_T^2)=0$ if $F(x,k_T^2)<0$
since $F(x,k_T^2)\geq 0$ according to the definition of the gluon
distribution.

    The chaotic solutions of Eq. (2.1) exist around $k_c\sim Q_s\sim 1~GeV$,
where perturbative calculations are barely available. However, more
lower $k_T$-range should be included, for example, we take
$k_0=0.1~GeV$. The region at $k_T^2<1~GeV^2$ is a complicate range,
where coexisting perturbative and non-perturbative effects. For
avoiding the difficulty in the infrared region, the evolution region
was divided into two parts at $Q_s=1~GeV$ in the previous work [9].
This is not a smooth treatment and it may deform the effects of the
chaotic solutions.

     Fortunately, many works have discussed the low
$Q^2$ transition region from the perturbative side [15]. They
incorporate in an effective non-perturbative corrections into the
evolution calculations. Considering the non-perturbative dynamics of QCD generate an effective gluon mass at
very low $Q^2$ region, and its existence is strongly supported by
QCD lattice simulations [16]. This dynamical gluon mass is
intrinsically related to an infrared finite strong coupling
constant. According to this idea, the suppressed strong coupling
constant can be used at low $Q^2$ and we take a following
restriction

$$\alpha_s\leq\alpha_{s,Max}\equiv B, \eqno(2.3)$$ the constant B
describes the non-perturbative corrections and we take $B=0.5$ as an
example. We will indicate that our results are insensitive to the
value of B in a reasonable range. Thus, we can expand our
perturbative calculation to $k_0=0.1~ GeV$.

    The $x$-dependence of $F(x,k_T^2)$ with different values
$k_T^2$ are illustrated in Fig. 1. It is surprise that one (thick)
line with $k^2_c=0.654~GeV^2$ approaches to a large positive value
at $x\rightarrow x_c=6\times 10^{-6}$, (this line has not been
reported in Ref. [9]); while all other lines with $k_T^2\neq k^2_c$
drop suddenly to zero. This result seems that the gluons in the
proton converge to a state with a critical momentum ($x_c,k_c)$.

    The $k_T^2$-dependence of $F(x,k_T^2)$ in Fig. 2 more clearly shows
the evolution of the gluon distribution from the saturated input to
the condensed state step by step. This result indicates that a lot
of gluons in the proton converge to a state at a critical momentum
($x_c,k_c)$, i.e. a typical gluon condensation.

    Figure 3 presents the value $x_c$ with different parameter
$\alpha_{s,Max}$ in Eq. (2.3). We find that the gluon condensation
still exists in a reasonable range of $\alpha_{s,Max}$.

    We recalculated Eq. (2.1) but the evolution region was
divided into perturbative and non-perturbative region and treated
separately as in Ref. 9. In this method, the evolution region has
two parts: region(A) 0 to $Q^2_s$ and region(B) $Q^2_s$ to $\infty$.
In region(B) the QCD evolution equation (2.1) is taken to evolute
and in region(A) the nonperturbative part of $F(x,k^2)$ is
identified as

$$F(x,\underline{k}^2)=C\underline{k}^2{\mathcal{F}}_{GBW}(x,\underline{k}^2),
~~at~x\leq x_0,~\underline{k}^2\leq Q^2_s,\eqno(2.4)$$ where the
parameter $C$ keeps the connection between two parts. The results
are shown in Figs. 4 and 5, where the dashed lines are proportional
to the GBW input according to Eq. (2.4). We find the peak at $x_c$
similar to Fig. 1, but the $k_T$-dependent structure of $F(x,k^2_T)$
near $x_c$ is completely distorted due to Eq.(2.4), which shadows
the evolution of the condensation at $k^2_T<k^2_c$. Note that if we
removed the dashed lines from Figs. 4 and 5, the results are
consistent with Figs. 1 and 2. Therefore, the treatment in
Eq. (2.4) hiders our understanding of the condensation solution.

       One can understand the above gluon condensation as follows.
As work [9] has pointed out that the derivative structure $\sim
\partial F(x,k_T^2)/\partial k_T^2$ and $\sim \partial
F^2(x,k_T^2)/\partial k_T^2$ in Eq. (2.1) add a perturbation on the
smooth curve $F(x,k_T^2)$ once $k_T$ crosses over $Q_s$. Thus, we
have a series of independent perturbations in a narrow $k_T$ domain
near $Q_s$ along evolution to smaller $x$. In the linear BFKL
equation, these perturbations are independent and their effects are
negligibly small. In this case the solutions keep the smooth curves
in ($x$, $k_T^2$) space. However, the nonlinear Eq. (2.1) may form
chaos near $Q_s$. The positive Lyapunov exponents of Eq. (2.1) in
Fig. 6 support this suggestion. Note that once chaos is produced,
the fast oscillations of the gluon density produce both the negative
and positive nonlinear corrections to $\Delta F(x,k^2_T)$ through
the derivative structure of Eq. (2.1). The former shadows the
grownup of the gluon density, while the later is the antishadowing
effect, and it increases festally the gluon density because it is a
strong positive feedback process. A maximum distribution $F(x\sim
x_c, k^2_T\sim k^2_c)$ in Fig. 1 will result a pair of closer and
more stronger positive and negative corrections at a next evolution
step, where the positive correction continually put $F(x,k_T^2)$
toward to a biggest value, while the negative one suppress all
remain distributions. Thus, we observed the gluons condensation at
($x_c,k^2_c$) due the extrusion of the shadowing and antishadowing
effects in the QCD evolution. Comparing with the CGC, Figs. 1 and 2
show a really gluon condensation in the gluon distribution.

   The peak value $F(x_c,k^2_c)$ is uncertain, although it is a big
value. According to the character of the condensate, the infinite
Bosons converge to a same point on the phase space, an ideal
$F(x,k^2_T)$ at $(x_c,k^2_c)$ is the delta-function. However, any
measurable distribution $F(x_c,k^2_c)$ has a width and the
corresponding peak value is finite, which depends sensitively on the
measurement conditions and even on the calculating precision.
Therefore, the precise value of $F(x_c,k^2_c)$ should be determined
by the experiments.

\newpage
\begin{center}
\section{The effects of the gluon condensation}
\end{center}

    The gluon condensation in the example of Sec. 2 produces the big corrections
to the normal parton distributions even by several orders of
magnitude. Such strong signals should appear in the experimental
data if the probe enters an enough lower $x$ range containing $x_c$.

    Unfortunately, we can not determine the value of $x_c$ in the
theory since several uncertainties. For example, the value of $x_c$
relates sensitively to the starting position $x_0$ of Eq. (2.1),
which is really unknown although we assumed $x_0=4\times 10^{-5}$ at
Sec. 2. A similar example is the BFKL equation. Usually we assume
that the BFKL equation starts work at $x\sim 10^{-3}-10^{-4}$,
however, the most PDF databases apply the DGLAP equation till to
$x<< 10^{-3}$. Besides, the uncertainties of the parameters in Eqs.
(2.1) and (2.2) also hider us to predict the value of $x_c$.
Therefore, we take the following program to make the estimations of
the gluon condensation effects: In this section we take the example
of Sec. 2 to study the effects of the gluon condensation, then we
transplant the results to astrophysics, where we may obtain the
information about the value of $x_c$. Finally, we predict the
signals of the gluon condensation in the future hadron colliders
using the determined $x_c$.

    The cross section of inclusive particle
production in high energy proton-proton collision is dominated by
the production of gluon mini-jet using the unintegrated gluon
distribution via [17,18]

$$\frac{d\sigma}{dk_T^2dy}=\frac{64N_c}{(N^2_c-1)k_T^2}\int_{0.1}^{100}q_T d
q_T\int_0^{2\pi}
d\phi\alpha_s(\Omega)\frac{F(x_1,\frac{1}{4}(k_T+q_T)^2)F(x_2,\frac{1}{4}
(k_T-q_T)^2)}{(k_T+q_T)^2(k_T-q_T)^2}, \eqno(3.1)$$where
$\Omega=Max\{k_T^2,(k_T+q_T)^2/4, (k_T-q_T)^2/4\}$; The longitudinal
momentum fractions of interacting gluons are fixed by kinematics:
$x_{1,2}=k_Te^{\pm y}/\sqrt{s}$; The distribution $F(x,k^2_T)$ is
taken from the results in Sec. 2 but they are multiplied by
$(1-x)^4$ for expanding to $x>x_0$.

    The rapidity distribution of the gluon-jets

$$\frac{d\sigma}{dy}=
\int_{x^2_cse^{\mp 2y}}^{100}d
k_T^2\frac{d\sigma}{dk_T^2dy},\eqno(3.2)$$where if $x^2_cse^{\mp
2y}<0.3~GeV^2$ we fixed it as $0.3~GeV^2$. The part of results at
different $\sqrt{s}$ are presented (solid lines) in Fig. 7. We find
that the large fluctuations arisen by the gluon condensation. The
gluon condensation effects begin work from $\sqrt{s_{GC}}\simeq
200~GeV$ in this example. The relation between $x_c$ and
$\sqrt{s_{GC}}$ is kinematically determined as follows. Note that
the gluon condensation plays a role if the contributions of the
gluon condensation peak local at $y_{Max}=\ln(\sqrt{s}/k_{T,Min})$,
i.e.,

    $$x_c=\frac{k_T}{\sqrt{s_{GC}}}e^{-y_{Max}}=\frac{k_Tk_{T,Min}}{s_{GC}}\simeq \frac{k^2_{T,Min}}{s_{GC}}.\eqno(3.3)$$
In the example Fig. 7, the resulting $\sqrt{s_{GC}}\simeq 200~GeV$.

    For the comparison, we plot the solutions removed the gluon
condensation by broken lines in Fig. 7, (i.e., the peak-like
distribution is removed from $F(x,k_T^2)$). The dashed lines are the
solutions using the GBW input but without the QCD evolution.
Comparing these lines, one can find the strong effects caused by the
gluon condensation in hadron collisions. Unfortunately, we never got
any repots about these effects till at the proton-proton collider with $E_{CM}=14~TeV$ in LHC. Therefore, we suggest that
$x_c<<6\times 10^{-6}$. In next section we try to determine the
value of $x_c$ using the possible signals of the gluon condensation
in astrophysics, where the energy scale of the proton-proton
interaction may be more larger then that in the accelerators.

\newpage
\begin{center}
\section{The gluon condensation and the GZK puzzle}
\end{center}

    Before 50 years, Greisen, Zatsepin and Kuz'min [13] predicted a drastic reduction
of the spectrum of cosmic rays around the energy of $E= (2\sim
5)\times 10^{19}~ eV$, since energy losses of the cosmic rays in the
cosmic microwave background radiation during their long propagation.
This is the GZK cutoff.

    The mean free path for photoproduction is calculated by
$\lambda_{\gamma p}=1/(N\sigma)$, where N is the number density of
blackbody photons and $\sigma(\gamma p\rightarrow
\pi^0p)\simeq100\mu b$ is the cross section at threshold. This leads
to $\lambda_{\gamma p}\simeq 10Mpc$. The Markarian galaxies are the
nearest possible UHECR-sources, which are residing at distances of
approximately $x\sim 100 Mpc$. The arrival probability of protons
through these distances with energies exceeding $10^{20} eV$ is only
$\sim e^{-x/\lambda_{\gamma p}}=10^{-4}-10^{-5}$. However, the
observations defy this result [19-21], where the recent Auger data
seem to diminish by steps only in one order of magnitude, but not by
an abrupt descend as above conceived. A big gap presents between
theory and experiments. Many ideas and different models are proposed
to understand the GZK puzzle even suspecting the Lorentz invariance
and the Standard Model, however, the true answer of the GZK puzzle
is still far from knowing.

    We noticed the following facts: since the flux of UHECRs is so low, direct
measurement of properties of UHECRs on the earth is impractical. One
must measurement is the extensive air shower on the earth, which is
created when cosmic ray enters the atmosphere.

The total cross section measured in the proton-proton collision is
generally defined as

$$\sigma=\frac{J}{n_{beam}},\eqno(4.1)$$where $J$ is the total number of measured interactions and
$n_{beam}=J_0/\sigma_0$ is the number of beam particles per unit
$\sigma_0$ of transverse area. Therefore, the detected UHECR flux on
the earth reads

$$J(E)=\frac{\sigma(\sqrt{s})}{\sigma_0}
J_0(E),\eqno(4.2)$$ where $J_0(E)$ is the primary flux of UHECRs;
$\sigma$ is the interaction cross section of the proton in the
UHECRs with the atmospheric proton.  Note that the GZK energy scale
$E\sim 2\times 10^{19}~eV$ corresponds to the total energy in the
center of mass (CM) frame $\sqrt{s}\sim 200~TeV$ using
$\sqrt{s}\simeq \sqrt{2m_NE}$; $E$ is the interaction energy in the
rest frame of the target proton. Obviously, such energy far exceed
the energy of existing particle accelerators. There are no any
reasons to indicate that the cross section $\sigma$ at $\sqrt{s}>
200 ~TeV$ still keeps the traditional estimation. We assume that the
value of $x_c$ is enough small, and a sudden increase of the
proton-proton cross section at such GZK-scale due to he gluon
condensation may fill the GZK suppression.

    Now let us to realize this idea. At first step, we calculate the
corrections of the gluon condensation to the cross section of
proton-proton collision in the example of Secs. 2-3, where
$x_c=6\times 10^{-6}$ is used. We define the rate

    $$R(\sqrt{s})\equiv \frac{\int dy\frac{d\sigma(\sqrt{s})}{d y}}
{\int dy\frac{d\sigma_{naive}(\sqrt{s})}{dy}}, \eqno(4.3)$$where
$\sigma_{naive}$ is the cross section without the QCD evolution
(dashed lines in Fig. 7). The rate $R$ represents the corrections of
the gluon condensation to the proton-proton cross section at
different scale $\sqrt{s}$. The results are shown in Fig. 8a.

    In the next step, we transplant the results with $x_c=6\times 10^{-6}$ to a
more small critical value $x^I_c\ll x_c$. For this sake, we need a
new set of $F(x,k^2_T)$ with $x_0\ll 4\times 10^{-5}$. For
simplicity, we take an indirect way to do them. We have pointed out
that the strong gluon condensation effects are dominated by the peak
distribution $F(x_c,k^2_c)$ at $x_c$. According to Eq. (3.3) we use
$$\frac {x_c}{x^I_c}=\frac{s^I_{GC}}{s_{GC}}\rightarrow\frac{s^I}{s}
\eqno(4.4)$$ to estimate the value of $x_c^I$ corresponding to
$s^I_{GC}$. The above last step assumes that this scale transform is
valid at $s>s_{GC}$. Thus, if the gluon condensation begins work at
the GZK scale $E=2\times 10^{-19}~eV$ (or
$\sqrt{s^I_{GC}}=200~TeV$), we should choose $x^I_c=6\times
10^{-12}$. Using this result and Fig. 8a we modify the proton-proton
cross section at $\sqrt{s^I}\geq 200~TeV$ as $\sigma(s)\rightarrow
\sigma^I(s^I)= \sigma(s^I)/R(s^I)$ in Eq. (4.2), where a new $s^I$
scale in Fig. 8a is used. Figure 9 shows the cosmic-ray energy
spectrum measured by the Auger collaboration [21]. The spectrum is
divided by $E^{-2.67}$. The open point and open star are the results
of the Auger data divided by R in Fig. 8a. The solid line in Fig. 9
is a smoothing result. It is surprise that the gluon condensation
may suddenly enhance the proton-proton cross section by almost four
orders of magnitude, they may fill the GZK suppression.

    The index in the power law $J\propto E^{-\gamma}$ in our results
is $\gamma\simeq 17$ at $2\times 10^{19}- 3\times 10^{19}~ eV$. It
is much larger than the power index $\gamma=2.67$ at $E<2\times
10^{-19}~eV$ and presents a sudden fall in the energy spectrum as
predicted by GZK cutoff.

    We consider another possible choice of $x_c$: the gluon condensation
starts from $\sqrt{s^{II}_{GC}}=80~TeV$, where is a position of the
ankle at $E=3.5\times 10^{18}~eV$. In this case, $x^{II}_c=4\times
10^{-11}$. The results using Fig. 8b are presented by the dashed
line in Fig. 9.

    The flux $J_0(E)$ can be estimated by the interaction length $L(E)$
using [22]

    $$J_0(E)\simeq \frac {1}{4\pi}L(E)\Phi(E),\eqno(4.5)$$where the local injection
spectrum $\Phi(E)$ has a power-law form of the hadron spectrum $\sim
E^{-2.67}$ in energy. We can not determine $J_0(E)$ since the
position of the UHECR-source is not fixed. However, the generally
expected proton interaction length quickly reduces a few orders of
magnitude at the GZK scale [22,23], and this is consistent with our
results in Fig. 9.

   The saturation and condensation origin from the BK and
Eq. (2.1), respectively. One can understand a big difference between
the starting points of these two evolution equations. The nonlinear
terms in the BK equation exclude the contributions of the gluon
recombination in the cross-channels [24]. These processes are
considered by Eq. (2.1) at more higher density of gluons, where the
correlations among gluons becomes stronger. However, the enhancement
of the gluon density with increasing $x$ at the saturation range is
very slow due to a big shadowing. Therefore, the starting point
$x_0$ of the evolution in Eq. (2.1) is much smaller than that in the
BK equation.

     We noted that the Auger collaboration reported [25] that the proton-proton cross section at
$\sqrt{s}=57~TeV$ is a normal value $\sim 505 ~mb$. This energy
scale is close to $\sqrt{s^{II}_{GC}}=80~TeV$. However, the result
is derived indirectly from the distribution of the depths of shower
maximum, its tail is sensitive to the cross section. We think that
the true shower shape originates from the condensate gluons,
therefore, it is different from the normal shower shape since the
coherence among the gluons at $x_c$. Therefore, we can not exclude a
strong proton-proton cross section at this energy scale.

\newpage
\begin{center}
\section{The gluon condensation at the future hadron colliders}
\end{center}

    The projected high energy proton-proton collisions
will probe deeply the very low $x$ domain, where we may observe the
gluon condensation. According to the GZK cutoff we have two possible
choices: (i) $\sqrt{s^I_{GC}}=200~TeV$, $x^I_c=6\times 10^{-12}$;
(ii) $\sqrt{s^{II}_{GC}}=80~TeV$, $x^{II}_c=4\times 10^{-11}$. We
give the rapidity distributions of the gluon jets at proton-proton
collision with $E_{MC}=100~TeV$ in Figs. 10a and 10b for these two
assumptions, where y-scale is re-plotted using
$y_{Max}=\ln(\sqrt{s^{I(II)}}/k_{Min})$. We find that in the
$x^{II}_c$ case, the gluon condensation effect is obvious.

    The fluctuation structure also appears in the transverse-momentum
distributions of the gluon jets

$$\frac{d\sigma}{d k_T}=2k_T\int_0^{y_{Max}}dy\frac{d\sigma}{dk_T^2dy}.\eqno(5.1)$$
The results are shown in Fig. 11, where the broken and dashed lines
are the results from removing the contributions of the gluon
condensation and the input distribution without QCD evolution,
respectively. The results show that the contributions of the gluon
condensation are constructed by many sub-jets. The strength of these
sub-jets is much higher then the normal distribution. The gluons
inside every sub-jet are dominated by the condensate gluons, they
have same energy-momentum. In particulary, these gluons are created
at a same collision time and have the same phase. Therefore, the
gluons in every sub-jet are monochromatic and coherent. We call the
phenomena in Figs. 10 and 11 as the gluon-jet bursts. Although our
estimations are rough, such extremely intense gluon field are the
ideal laboratory studying QCD at the extreme conditions. We should
pay attention to the big effects of the gluon-jet bursts when
planning the next high energy hadron colliders and the detectors.

    The nuclear target may increase the value of $x_c$ since the
nonlinear corrections need to be multiplied by $0.5A^{1/3}$ in a
nucleus-nucleus collider [26], where the factor 0.5 is from the
nuclear geometric corrections. We take $Pb-Pb$ collider as the
example, the numeric solutions of Eq. (2.1) show that

$$\frac{x_c}{x_{c;Pb-Pb}}\simeq\frac{x_c^{I(II)}}{x^{I(II)}_{c;Pb-Pb}}.
\eqno(5.2)$$ We get $x_{c;Pb-Pb}^I=2\times 10^{-11}$ and
$x_{c;Pb-Pb}^{II}=10^{-10}$. Using Eq. (4.4) we present our
predictions in Figs. 12-15.

    E. Fermi predicted jokingly that a maximum accelerator will around the
equator. However, there is an applicable maximum energy for the
hadron-hadron collider.
At high energy (or at small $x$), the total cross section of the collision is responsible for the gluon distributions
in the beam nucleons. The gluon condensation implies that the gluons with $x<x_c$ converge to a
critic state at $x=x_c$, which leads to $F(x<x_c,k^2_T)=0$. This prediction should be presented
in the measurable cross section $\sigma_{p-p}$.
Note that for a
given collision energy, only the partons in a certain kinematic
range are effectively used due to the kinematic restriction.
We image that the condensate peak begins work at $\sqrt{s_{GC}}$.
As we have shown in Figs. 7 and 8, it rises a sudden big increase of
the proton-proton cross sections, and this effect expands till $\sqrt{s_{Max
}}$.  On the other hand, $F(x<x_c,k^2_T)$ will dominate the parton interacting range if
$\sqrt{s}>\sqrt{s_{Max}}$. The results in Fig. 7 show that
the gluon contributions to the hadron collider almost disappear when
the position of the condensation peak approaches to the rapidity center
$y=0$. The last three diagrams in Fig. 7 present this situation,
where the missing part of the rapidity distribution corresponds to
the disappearance of gluons at $x<x_c$ in Fig. 1. One can estimate
the corresponding energy scale $\sqrt{s_{Max}}$ using Figs. 7 and 8. We find that
$\sqrt{s^I_{Max}}\simeq 10^{6}~TeV$, or $\sqrt{s^{II}_{Max}}\simeq 10^{5}~TeV$
for our two assumptions. Beyond this energy
scale, $\sigma_{pp}$ is almost small, where the remaining small
contributions are from the quarks and Abelian gluons [9].  It
implies that a proton beam becomes "transparent", therefor, the high
energy collider is inefficient at $\sqrt{s}>\sqrt{s_{Max}}$. We call
$\sqrt{s_{Max}}$ as the maximum applicable energy of the proton-proton collider.

    A purpose of the high energy collider is to convert the kinetic energy of
the beam nucleons into the creating new particles. A big cross section $\sigma_{p-p}$
implies a high rate of this conversion. Therefore, $\sqrt{s}=100-10^6~TeV$ is a golden
energy range for the proton-proton collider.

\newpage
\begin{center}
\section{Discussions}
\end{center}

    The equation (2.1) is based on the leading QCD approximation, where
the higher order corrections are neglected. An important question
is: will disappear the chaos effects in the evolution equation after
considering higher order corrections? We answer this question from
two different aspects.

    (i) As we have pointed out that chaos in the modified BFKL equation
origins from the special singularity of the nonlinear evolution
kernel. From the experiences in the study of the BFKL equation, the
higher order QCD corrections can not remove this primary singularity
[27]. Let us assume that Eq.(2.1) is modified as following form if
considering the higher order corrections

$$-x\frac{\partial F(x,k_T^2)}{\partial x}$$
$$=\frac{3\alpha_{s}k_T^2}{\pi}\int_{k^2_0}^{\infty} \frac{d k
'^2_T}{k'^2_T}\left\{\left[\frac{F(x,k'^2_T)-F(x,k_T^2)} {\vert
k'^2_T-k_T^2\vert}+\frac{F(x,k_T^2)}{\sqrt{k_T^4+4k'^4_T}}
\right]\left[[1+A (k'^2_T,k_T^2)\right]\right\}$$
$$-\frac{81}{16}\frac{\alpha_s^2}{\pi R^2_N}\int_{k^2_0}^{\infty} \frac{d k
'^2_T}{k'^2_T}\left\{\left[\frac{k_T^2F^2(x,k'^2_T)-k'^2_TF^2(x,k_T^2)}
{k'^2_T\vert
k'^2_T-k_T^2\vert}+\frac{F^2(x,k_T^2)}{\sqrt{k_T^4+4k'^4_T}}\right]\left[
1+B (k'^2_T,k_T^2)\right] \right\}. \eqno(6.1)$$ One can image that
the contributions from $A(k'^2_T,k_T^2)$ and $B(k'^2_T,k_T^2)$
either are the smooth function of $k'^2_T$ and $k_T^2$, or have the
extra singular structure. In the former case, we take an
approximation: A and B are almost constant and

$$-x\frac{\partial F(x,k_T^2)}{\partial x}$$
$$=\frac{3\alpha_{s}k_T^2}{\pi}\int_{k^2_0}^{\infty} \frac{d k
'^2_T}{k'^2_T}\left\{\left[\frac{F(x,k'^2_T)-F(x,k_T^2)} {\vert
k'^2_T-k_T^2\vert}+\frac{F(x,k_T^2)}{\sqrt{k_T^4+4k'^4_T}}
\right]\beta \right\}$$
$$-\frac{81}{16}\frac{\alpha_s^2}{\pi R^2_N}\int_{k^2_0}^{\infty} \frac{d k
'^2_T}{k'^2_T}\left\{\left[\frac{k_T^2F^2(x,k'^2_T)-k'^2_TF^2(x,k_T^2)}
{k'^2_T\vert
k'^2_T-k_T^2\vert}+\frac{F^2(x,k_T^2)}{\sqrt{k_T^4+4k'^4_T}}\right]
[ 1-\beta] \right\}. \eqno(6.2)$$ We give the predicted value $x_c$
with different values of $\beta$ in Fig. 16. One can find that the
gluon condensation solution is insensitive to the parameter $\beta$
in its reasonable range.

    In the second case, Eq. (2.1) may have the multi-chaos solution.
For example, we take the Fadin-Lipatov (KL) model [28] as the input
to study Eq. (2.1), i.e.,

$$F(x_0,k_T^2)=\left\{\begin{array}{ll}
f_0 k_T^2 ~if~ k_T^2<Q^2_s\\
f_0Q^2_s ~if~ k_T^2> Q^2_s \end{array}\right. \eqno(6.3)$$ The
solution shows two positive peaks in Fig. 17, which correspond to
two maximum values of Lyapunov exponents in Fig. 18. One of them
arris from a non-smooth connection at $k_T^2=Q^2_s$ in Eq. (6.3).
However Fig. 19 shows that these two chaos lead to the gluon
condensation at a critic value $x_c$ because the competition among
several positive feedback processes. This conclusion has a general
mean: if existing the multi-singular structure from the higher order
corrections, the corresponding nonlinear evolution equation still
has the gluon condensation.

    (ii) We discuss the approximation solution of Eq. (2.1) from the
view point of the chaos theory. It is well known that some of
chaotic attractors are unstable. A slight fluctuation of a parameter
may drive the system out of chaos. However, it has been proofed that
some dynamical systems can exhibit robust chaos [29]. A chaotic
attractor is said to be robust if, for its parameter values, there
exist a neighborhood in the parameter space with absence of periodic
negative Lyapunov exponents. Robustness implies that the chaotic
behavior cannot be destroyed by arbitrarily small perturbations of
the system parameters. The structure of the Lyapunov exponents in
Figs. 6 and 18 show absence of any negative values around $k_T^2\sim
1 ~GeV^2$, and the maximum value of $\lambda$ is enough larger
$\lambda\gg 1$. This means that chaos in Eq. (2.1) is robust.
Therefore, we expect that chaos and its effects still exist even
considering the higher order corrections.

    The above analysis tell us that the gluon condensed effects origin from the singular nonlinear
evolution kernel, which is a general structure in the logarithmic
($1/x$) resummation. Now we point out that the gluon condensation is
a nature result of the momentum conservation. We call the positive
corrections of the nonlinear terms in a QCD evolution equation as
the antishadowing, which is the compensation to the shadowing effect
due to the momentum conservation [30]. There are two different
antishadowing effects: one was presented in a modified DGLAP
equation [31] and a modified BK equation [24], where the
antishadowing effect compensates the lost momentum in shadowing.
Since in these examples the shadowing is smaller and the increasing
gluons distribute in a definite kinematic range, such antishadowing
effect is weaker and it consists with the observed EMC effects [32].
On the other hand, in the gluon condensation, a lot of gluons
compensate the disappearing gluons at $x<x_c$, and they accumulate
at a same critic momentum. In consequence, a sharp peak is added on
the gluon momentum distribution and it creates a series of strong
effects. Therefore, the gluon condensation is an inevitable result
due to the momentum conservation for compensating the lost momenta
in the blocking QCD evolution.

\newpage
\begin{center}
\section{Summary}
\end{center}

    A QCD evolution equation at small $x$ should sum the
contributions of the gluon random evolution on the
transverse-momentum space. The evolution kernels of this equation
have singular structure even in the nonlinear kernels. A standard
regularization technic is to sum the contributions of the virtual
diagrams according to the unitary theory. The resulting evolution
kernels have approximately the derivation form with $k_T$. A
modified nonlinear BFKL equation Eq. (2.1) is a such example.

    Equation (2.1) has the robust chaotic solution arising from its
nonlinear singular structure if the input distribution has an
obvious deformation likes the saturation form around $k_T\sim Q_s$.

    In this work we present that the dramatic chaotic oscillations produce the strong shadowing and
antishadowing effects, they converge gluons at $x<x_c$ to a state
with a critical momentum $(x_c,k_c)$. This is the gluon condensation
and the blocking QCD evolution.

    The sharp peak in the momentum distributions caused by the gluon condensation
implies a large enhancement of the cross section in hadron-hadron
collision. We examine that the sudden increase of the proton-proton
cross section by several orders of magnitude may fill the GZK
suppression. Using this result we extract the critic parameter
$x_c$. Then we predict the possible observations of the gluon
condensation effects in the future hadron colliders. We predict a
maximum applicable energy of the hadron collider due the blocking
QCD evolution of the gluons. We find that the gluon condensation
leads to the big fluctuations of the gluon jets in its rapidity and
transverse-momentum distributions at a ultra high energy range. The
gluons in every sub-jet are monochromatic and coherent, and we call
them as the gluon-jet bursts. Such extremely intense gluon field
caused by the gluon condensation is an ideal laboratory to study QCD
at the extreme-conditions. We should pay attention to the big
effects of the gluon-jet bursts when planning the next high energy
hadron colliders and the detectors.

\noindent {\bf Acknowledgments}: We thank F. Wang for useful
discussions. We would also like to thank X.R. Chen for organizing a
meeting in IMP, where we discussed the gluon condensation. One of us
(J.S.L) thanks H.K. Dai, J.H. Ruan and R. Wang for their help.

\end{document}